\documentclass{ws-procs975x65}

\begin{document}

\title{Effects of BBN on population III stars}

\author{F. Iocco}

\address{Complesso Universitario di M.te S.Angelo, via Cinthia,
80126 Napoli, Italy.\\E-mail: iocco@na.infn.it}

\maketitle

\abstracts{
The presence (or absence) of CNO elements in the primordial gas
determines different behaviours in population III stars formation and
evolution: we therefore present an analysis of the main channels for
the synthesis of these elements in BBN in order to understand, within
a reliable interval, their abundance in the primordial material.}

\section{Introduction}
\noindent
{Recent studies on population III (popIII) showed how interesting this
pregalactic generation of stars is for our overall understanding of
cosmology: their contribution to the first stucture formation and to
reionization makes them an extremely promising topic, (see for example
\cite{cen}). In light of the huge number of works published in the
last years, (see \cite{bromm} and references therein), it's likely to
picture the population III as a self-killing generation of very
massive stars, which evolve fastly and eventually explode as Pair
Instability Supernovae, ejecting in the outer space the metals they
have produced during their life, (see for example \cite{SaFe}). The
initial mass function (IMF) of these stars is determined from the
absence of metals in the material from which they form out: there is a
threshold in the metallicity of the gas above which the IMF shows the
usual Salpeter form, and below which the IMF is the top-heavy one
shown from \cite{Nakamura}. The paucity of metals affects also the
evolution of popIII: as widely known stars more massive than 3-4
M$_{\odot }$ burn hydrogen via CNO cycle, so that the scarcity of CNO
catalyzers in popIII forces the main sequence through an "unusual" pp
burning of hydrogen, resulting in an expansion and extreme heating of
the core \cite{Straniero}. This phenomenon shows a threshold in CNO
elements abundance, above which the star experience a "normal" main
sequence. In light of all this it would be extremely important to find
out with reliable precision the amount of metals in the primordial
gas.

\section{Metal nucleosynthesis in BBN}
\noindent
The successes of BBN predictions, obtained comparing the theoretical
estimates of light elements with the observed abudances, show the deep
knowledge of nucleosynthesis phenomenon we have today -at least for
what deals with light elements \cite{noi}. In particular the updating
of the cross-section data and the knowledge of the main synthesis
channels is essential in order to estimate the primordial elemental
abundances and their uncertainties. Such a precision in the light
nuclide predictions is not yet available for the heavier CNO
nuuclides, nor in the connections between light and heavy elements.
Historycally this is due to the very low final abundance of these
elements which does not justify an effort to determine them, as long
as it is impossible to compare the predictions with observational
data. In light of the importance of metals (and more specifically of
CNO elements) in POPIII stars theory we have performed an analysis of
the BBN network toward carbon, nytrogen and oxygen, identifying the
main reaction channels and updating, where possible, the cross-section
data with newest ones. Our study showed that the main channel for
carbon production in stars, namely the $2^{4}He(\alpha ,\gamma
)^{12}C$ which provides a direct connection between light and heavy
elements, is suppressed in BBN as a consequence of the very low
density of the plasma; we have identified as main channels for carbon
synthesis the {$^{7}Li(\alpha ,\gamma )^{11}B(p,\gamma )^{12}C$},
{$^{7}Li(\alpha ,\gamma )^{11}B(d,n )^{12}C$} and {$^{7}Be(\alpha
,\gamma )^{11}C$}, together with the back reaction $^{11}C(n,\alpha
)2^{4}He$ and $^{11}B(p,\alpha )2^{4}He$; heavier elements build up
through progressive proton, deuteron and neutron captures upon carbon
nuclei. These results, obtained by computing the contributions of the
different reactions to the Boltzmann equation which describes the
elemental abundance evolution, show that the heavy elements synthesis
in BBN is strongly related to the intermediate element (such as $Li$,
$Be$ and $B$) one, thus emphasizing the necesssity of a complete
revision of the "intermediate" and "upper" part of the network. We
have also noted that most of the reactions involving isotopes of
hydrogen and helium but $H$ and $^{4}He$ are neglected in the existing
code -at least for what deals with the "metal network"; we have added
a first set of new reactions to study how they affect the
heavy-nuclide synthesis, and results will be presented in a future
paper.

\end{document}